\documentclass[11pt]{article}

\usepackage[letterpaper,margin=1in]{geometry} 
\usepackage{graphicx}
\usepackage{amsmath}
\usepackage{url}
\usepackage{hyperref}
\usepackage{booktabs}
\usepackage{threeparttable}
\usepackage{float}
\usepackage{caption}
\usepackage{subcaption}
\usepackage[backend=biber,style=apa,natbib=true]{biblatex}
\DeclareLanguageMapping{american}{american-apa}
\title{Do High-Premium Fields Buffer Labor Market Shocks? Evidence from India}
\author{Jheelum Sarkar\footnote{Department of Economics, American University, Washington DC-20016. Email:js8622a@american.edu}}
\date{May 2026}

\begin{document}

\maketitle
 \begin{abstract}
\noindent Do high-return fields of study provide greater protection in labor market during crises? I construct pre-pandemic premia for major technical fields in India and examine whether workers in higher field-premium fields experience resilient labor market outcomes during COVID-19. Using a difference-in-difference with continuous treatment design, I find that field-premium advantages did not emerge immediately at the onset of the pandemic but through gradual adjustment during later phases. 


\end{abstract}
\section{Introduction}
In many emerging economies, expansion of higher education has outpaced job growth \citep{yamada2018college}. For example, less than 4\% Indian college graduates secure white-collar jobs within a year from their graduation \citep{StateOfWorkingIndia2026}. It is widely known that graduates with different majors experience substantial differences in earnings (e.g., \cite{10.1257/aer.104.5.387, Walker_differences_2011, Altonji_the_2016, altonji2025returns}). These disparities raise an important question: \textit{do higher-return fields yield more resilient labor market outcomes at times of crisis?}\\

Differences in earnings across fields (i.e., field-of-study premia) reflect variation in skills, adaptability, and labor demand which become especially relevant in the labor market during economic disruptions. Fields differ in the extent to which they provide general versus field-specific skills \citep{LeightonSpeer2020}. For example, business degrees emphasize on general skills such as communication and organization while nursing provides more specialized skills \citep{NBERw29605, LeightonSpeer2020}. Fields also differ in occupational sorting. Individuals trained in math- and writing-intensive fields are more likely to enter higher-paying occupations \citep{Weiss_why_2024}. Technical fields such as engineering and medical sciences depict exhibit higher and stable income growth while liberal arts tend to have relatively lower-return trajectories \citep{Andrews_the_2024}. Thus, field-of-study premia can play an important role in shaping labor market resilience.\\

Existing studies on labor market adjustment during and after COVID-19 have primarily focused on occupational characteristics such as remote work and task content \citep{Adrjan_working_2025, Blanas_covid-induced_2023}. Recent evidence from the United States show that STEM employment was more resilient than non-STEM employment during the pandemic \citep{Davis_stem_2026}. However, it remains unclear whether this occupation-specific resilience could be driven by their fields of study. In this paper, I shift the focus from occupations to fields of specialization and examine whether high-return fields yield more resilient labor market outcomes during three phases of COVID-19 pandemic. This perspective is important because expectations about how employment and earnings evolve during downturns can influence an individual's ex ante field-of-study choices. I focus on India, an emerging economy with one of the largest youth populations in the world, where job mismatch is predominant \citep{Mehrotra2014}.\\

This paper contributes to two bodies of work. First, it relates to the literature on  labor market adjustments during economic downturns \citep{Altonji_cashier_2016, Kahn_the_2010, Oreopoulos_the_2012, Davis_stem_2026, Choi_the_2020}. Existing studies show that workers in higher-return fields are less affected than those in lower-return fields during recessions \citep{ Altonji_cashier_2016}. I provide new evidence that this pattern does not emerge immediately when the pandemic hit. Workers in high-premium fields experienced 1.1\%-1.2\% more working hours during the later phases of COVID-19. It was only in the last phase when individuals with specialization in high-premium fields witnessed 2.2\% more monthly income. By doing so, I show that resilience unfolds over the course of a crisis. I further extend this literature by providing evidence from an emerging economy where youth unemployment is as much as 40\% \citep{StateOfWorkingIndia2026}. Second, this paper also contributes to the literature on degree choice and labor market returns \citep{10.1257/aer.104.5.387, Altonji_the_2016, Long_do_2015}. While existing studies estimate which majors yield more returns, better career trajectories and skills, I examine whether the high-return fields provide greater resilience during adverse shocks.\\

\section{Data and Research Method}\label{data-method}
\subsection{Data}
I utilize individual-level repeated cross-sectional data from each of the five rounds of Periodic Labor Force Survey (PLFS), starting from July 2017—June 2018 till July 2021—June 2022. PLFS is nationally representative labor survey conducted by the National Statistical Office (NSO) of India. 
It provides detailed information on key labor market indicators and socioeconomic indicators of each surveyed individual. Data analysis requires using population weights provided in unit-record files.\footnote{MLTS is the NSO multiplier (stored with two decimal places). NSS is the count of first-stage sampling units (FSUs) surveyed in a given sector × state × stratum × substratum for a single sub-sample; NSC is the corresponding count for the combined sub-samples. Following NSO guidance, the combined person-level weight is MLTS/100 if NSS = NSC, and MLTS/200 otherwise} I focus on individuals with college degrees and above.\\
Table~\ref{tab:summary_stats} reports the summary statistics of continuous variables from the entire sample during 2017-2024. 

\begin{table}[htbp]
\centering
\begin{threeparttable}
\caption{Summary Statistics for Continuous Variables}
\label{tab:summary_stats}

\begin{tabular}{lccccc}
\hline\hline
Variable & Obs. & Mean & Std. Dev. & Min & Max \\
\hline
Monthly regular earnings (USD) & 203{,}340 & 101.91 & 205.19 & 0 & 22{,}000 \\
Self-employment earnings (USD) & 203{,}340 & 30.06 & 188.50 & 0 & 66{,}000 \\
Monthly casual wage earnings (USD) & 203{,}340 & 0.89 & 9.52 & 0 & 440 \\
Monthly earnings (USD) & 203{,}340 & 131.96 & 267.42 & 0 & 66{,}000 \\
Weekly working hours & 203{,}340 & 28.45 & 28.16 & 0 & 140 \\
\hline
Age (Years) & 203{,}340 & 35.50 & 13.03 & 17 & 79 \\
Potential experience (Years) & 203{,}340 & 13.79 & 12.95 & 0 & 68 \\
\hline\hline
\end{tabular}

\begin{tablenotes}
\footnotesize
\item \textit{Notes:} The sample includes 203,340 graduate and post-graduate individuals aged less than 80 years. Based on the Mincer equation, experience is computed as current age minus years of formal education minus 6, with negative values set to zero. Original earnings are reported in Indian rupees and converted into U.S. dollars using the prevailing exchange rate (1 INR $\approx$ 0.011 USD).
\end{tablenotes}
\end{threeparttable}
\end{table}

\subsection{Technical Field Premium}
To construct field-level variation, I group individuals into six mutually exclusive categories based on their technical education: (i) no technical education, (ii) agriculture, (iii) engineering/technology, (iv) medicine, (v) crafts, and (vi) other technical subjects. The key limitation is that PLFS does not provide details about \textit{``other technical subjects"} and fields covered under\textit{``no technical education"}. Note that I restrict the sample to individuals with graduate and post-graduate degrees to ensure that the comparisons are made with similar levels. Using the pre-pandemic subsample (2017--2019), I construct technical field premia by estimating the following specification:

\begin{equation}\label{eq1}
\ln w_{idt} = \alpha + \sum_{f} \delta_f \mathbf{1}\{field_i = f\} + X_{idt}'\gamma + \mu_d + \mu_t + \vartheta_{idt},
\end{equation}
where \(w_{idt}\) denotes monthly earnings of individual \(i\) in district \(d\) and year \(t\). \(X_{idt}\) includes age, age squared, experience, experience squared, years of education, and indicators for sex, marital status, religion, and social group. $\mu_d$ and $\mu_t$ are district and year fixed effects. The reference category is graduates and post-graduates without technical education. $\vartheta_{idt}$ is the error term.

The estimated coefficients \(\delta_f\) capture reduced-form earning differentials of each technical field compared to non-technical fields of study in higher education. The estimates are shown in Table~\ref{tab:field_premia}. Next, I standardize these field premium values so that it has mean zero and standard deviation one. Then, I merge this measure into the 2017-2022 sample by assigning each individual their field-level premium values.

\begin{table}[htbp]\centering
\caption{Pre-Pandemic Technical Field Premia}
\label{tab:field_premia}
\begin{threeparttable}
\begin{tabular}{lc}
\toprule
Technical field & Log earnings premium \\
\midrule
Agriculture               & 0.294*** \\
                          & (0.074)  \\
Engineering/technology    & 0.344*** \\
                          & (0.016)  \\
Medicine                  & 0.381*** \\
                          & (0.032)  \\
Crafts                    & 0.144    \\
                          & (0.118)  \\
Other technical subjects  & 0.198*** \\
                          & (0.020)  \\
\midrule
Reference category        & No technical education \\
\bottomrule
\end{tabular}
\begin{tablenotes}
\footnotesize
\item Estimated coefficients of $\delta_f$ using eq~\ref{eq1}. The omitted category consists of graduate and post-graduate individuals with non technical background. Robust standard errors are reported in parentheses. *** $p<0.01$, ** $p<0.05$, * $p<0.1$.
\end{tablenotes}
\end{threeparttable}
\end{table}

\subsection{Empirical Strategy}
Following Callaway et al. (\citeyear{Callaway_difference-in-differences_2024}), I implement a difference-in-differences framework with continuous treatment. With experiences from the initial pandemic wave, adaptive capacity is likely to improve in  subsequent pandemic phases. This, in turn, may influence labor market resilience of individuals across different fields of higher education. To capture this, I split the entire COVID-19 period into three phases based on calendar time. I define $Phase_{1}$ corresponds to 2020 (pandemic onset and nationwide lockdown), $Phase_{2}$ to 2021 (second wave), and $Phase_{3}$ to early 2022 (January--March) in the following specification:
\begin{equation} 
y_{idt} = \alpha + \sum_{k=1}^{3} \beta_k \left( Premium_i \times \mathbf{1}\{\text{Phase}_t = k\} \right) + X_{idt}^{\prime}\theta + \eta_d + \eta_t + \varepsilon_{idt}
\label{eq:main-eq}
\end{equation}

In the above specification, $y_{idt}$ measures labor market outcomes, namely, weekly working hours and monthly earnings  of individual $i$ in district $d$ at time $t$. I log transform monthly earnings due to its positive skewness (figure~\ref{fig:fig_summ1}). To capture proportionate changes in hours worked
instead of just levels, I have also log transformed weekly working hour variable. $Premium_i$ is a continuous measure of pre-pandemic field-of-study premia (Table~\ref{tab:field_premia}) assigned to each individual based on their field of education. The reference category is the pre-pandemic period (2017--2019).\\ 
$\mathbf{X}_{idt}$ is a vector of individual characteristics, including age, experience, sex, marital status, education levels, religion, and social group. $\eta_d$ and $\eta_t$ denote district and time fixed effects, respectively. $\varepsilon_{idt}$ is idiosyncratic error term. \\
The coefficients of interest are $\beta_k$s which capture the relationship between labor market outcomes and field premia during each phase $k$ relative to the pre-pandemic period. 

\section{Results}\label{results}
\subsection{Parallel Trend Test}
I first examine the parallel trend assumption: in absence of the pandemic, differences in labor market outcomes across individuals with different field premia would remain constant over time. To evaluate this, I use the following dynamic difference-in-differences specification:
\begin{equation}
y_{idt} = \alpha + \sum_{\substack{k \neq -1}} \gamma_k \left( Premium_i \times \mathbf{1}\{\text{Event}_t = k\} \right) + X_{idt}'\lambda + \tau_d + \tau_t + \varepsilon_{idt}
\label{eq:event-study}
\end{equation}
where $\mathbf{1}(\text{Event}=k)$ is a dummy that takes value 1 if the observation is $k$ years since the pandemic hit or first wave (i.e., January 2020--December 2020). Here, the base period is 2019 (i.e., $t$=-1), that is the year before the pandemic hit.  $\tau_d$ and $\tau_t$ are  district and year fixed effects. $e_{idt}$ is the error term. If $\gamma_k$`s are statistically insignificant for each pre-pandemic years, it supports the parallel trend assumption. That is, there is no evidence of differential trends in outcomes before the shock hits.
\begin{figure}[H]
    \centering
    \includegraphics[width=0.7\textwidth]{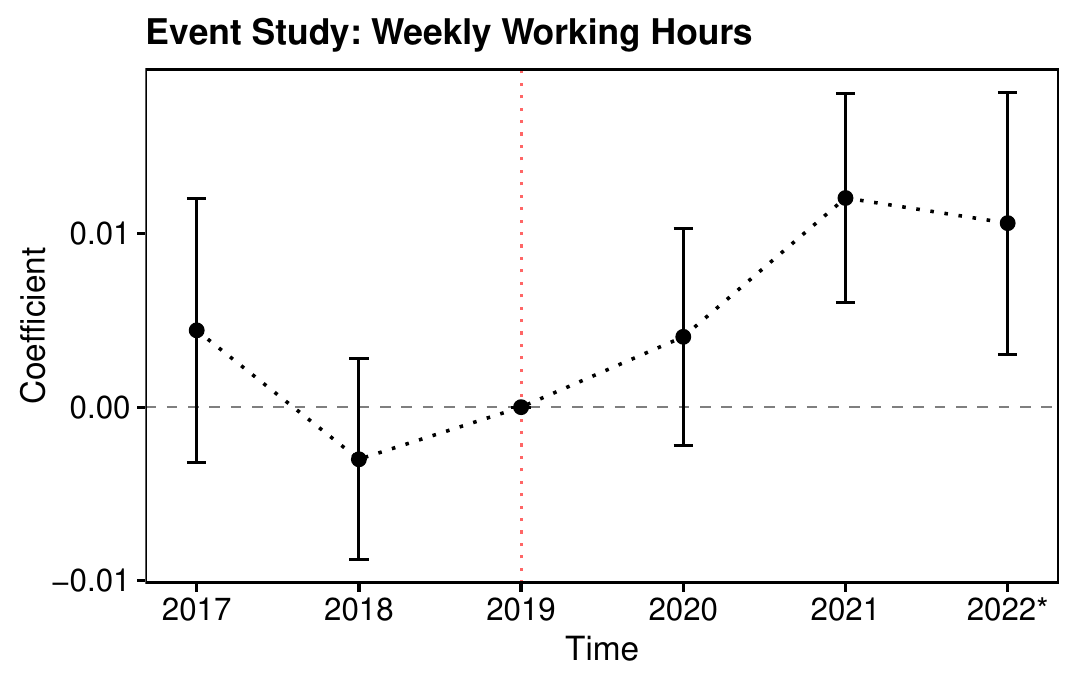}
        \caption{\textit{Dynamic difference-in-differences results for log of weekly working hours, with 2019 as reference period. Notes: Points are interaction-term coefficients; bars are 95\% confidence intervals. Here, 2022 includes the first two quarters of January-March and April-June.}}
        \label{fig:fig1}
    \end{figure}
Figures~\ref{fig:fig1} and~\ref{fig:fig2} report the dynamic difference-in-differences estimates for weekly working hours and monthly earnings. The estimated coefficients are close to zero and statistically insignificant in the two pre-pandemic years which supports the parallel trends assumption. It suggests that there is no evidence of differential pre-trends in labor market outcomes by field premium. The differences begin to emerge after the onset of COVID-19. During Phase 1 in 2020, there is no statistically significant differential in weekly working hours and monthly earnings. In later phases, however, workers in higher-premium fields experience stronger labor-market outcomes. A one-standard-deviation higher field premium is associated with about 1.2\% higher weekly hours in phase 2 and 1.1\% higher weekly hours in phase 3. However, the earning differential across fields emerge only in the last phase when a one-standard-deviation higher field premium is associated with about 2.5 \% higher monthly earnings relative to the pre-pandemic reference year. These patterns suggest that labor market resilience did not appear immediately among high premium fields when the pandemic hit but evolve gradually during the later phases, first through working hours and thereafter through income. 

\begin{figure}[H]
    \centering
    \includegraphics[width=0.7\textwidth]{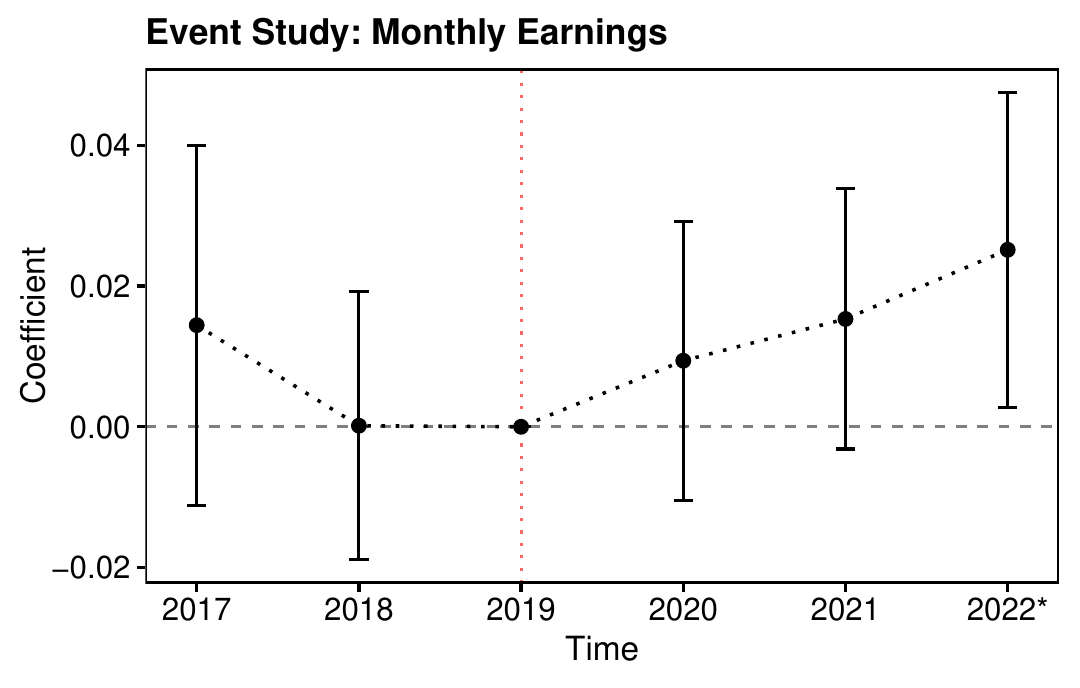}
    \caption{\textit{Dynamic difference-in-differences results for log monthly earnings, with 2019 as the reference period. Notes: Points are interaction-term coefficients and bars show 95\% confidence intervals. Here, 2022 includes the first two quarters of January--March and April--June.}}
    \label{fig:fig2}
\end{figure}

\subsection{Key Results}
Table~\ref{tab:main_results} presents the main regression results from equation \eqref{eq:main-eq} for log of weekly working hours (Panel A) and monthly earnings (Panel B). Column (1) shows results without controls and Column (2) includes controls.\\

Panel A shows that there is insignificant association between field premia and weekly working hours (Col.(1)~(2), Panel A, Table~\ref{tab:main_results}). But the relationship becomes stronger in the subsequent phases. A one-standard-deviation increase in field premium is associated with 1.2\% and 1.1\% more working hours per week during second and third phases respectively (Col.~(2), Panel A  Table~\ref{tab:main_results}). Panel B reports a similar but delayed positive association between monthly earnings and field premia. During both phase 1 and phase 2, earnings did not significantly change for every one-standard-deviation increase in field premia. The earnings differential emerges in Phase 3, where a one-standard-deviation higher field premium is associated with approximately 2.2 \% higher monthly earnings relative to the pre-pandemic period. This suggests that differences across fields first appeared through labor supply adjustment and later translated into earnings differences.\\ 

Table~\ref{tab:covid_results} reports pooled difference-in-difference estimate. I combined all phases of the pandemic into a single time dummy, ${Covid}$ which takes value 1 for all phases of COVID-19 and 0 for years prior the pandemic. Panel A shows that a one-standard-deviation increase in field premium is associated with 0.9-1.0 \% higher weekly working hours (Col.~(1)-(2), Panel A  Table~\ref{tab:covid_results}). Panel B also shows similar results where a one-standard-deviation increase in field premium is associated with 1.3 \% higher monthly earnings (Col.~(2), Panel B  Table~\ref{tab:covid_results}). \\

Thus, the findings indicate that high premium fields were associated with greater labor-market resilience during the pandemic. But the adjustment was not immediate: differences in working hours became visible in Phase 2, while earnings differences emerged only by Phase 3.

\begin{table}[h]\centering
\footnotesize
\caption{Field Premium and Labor Market Outcomes Across Pandemic Phases}
\label{tab:main_results}
\begin{threeparttable}
\begin{tabular}{lcc}
\toprule
 & (1) & (2) \\
 & No controls & With controls \\
\midrule
\multicolumn{3}{l}{\textbf{Panel A: Log Weekly Working Hours}} \\
\midrule
Field premium $\times$ Phase 1 & 0.005* & 0.004 \\
                               & (0.003) & (0.003) \\
Field premium $\times$ Phase 2 & 0.013*** & 0.012*** \\
                               & (0.002) & (0.002) \\
Field premium $\times$ Phase 3 & 0.012*** & 0.011*** \\
                               & (0.003) & (0.003) \\
\hline
Observations & 107,266 & 107,266 \\
R-Squared & 0.108 & 0.147 \\
\midrule
\multicolumn{3}{l}{\textbf{Panel B: Log Monthly Earnings}} \\
\midrule
Field premium $\times$ Phase 1 & 0.041 & 0.007 \\
                               & (0.080) & (0.008) \\
Field premium $\times$ Phase 2 & 0.070 & 0.012 \\
                               & (0.072) & (0.008) \\
Field premium $\times$ Phase 3 & 0.113 & 0.022** \\
                               & (0.096) & (0.010) \\
\hline
Observations & 99,137 & 99,137 \\
R-Squared & 0.192 & 0.321 \\
\midrule
District FE & Yes & Yes \\
Year FE & Yes & Yes \\
Controls & No & Yes \\
\bottomrule
\end{tabular}
\begin{tablenotes}
\footnotesize
\item Main results using eq.~(\ref{eq:main-eq}). The omitted category is the pre-pandemic period (2017--2019). Phase 1 corresponds to 2020, Phase 2 to 2021, and Phase 3 to January--March 2022. All specifications include district and year fixed effects. Robust standard errors are reported in parentheses. *** $p<0.01$, ** $p<0.05$, * $p<0.1$.
\end{tablenotes}
\end{threeparttable}
\end{table}

\begin{table}[h]\centering
\footnotesize
\caption{Field Premium and Labor Market Outcomes During the Pandemic Period}
\label{tab:covid_results}
\begin{threeparttable}
\begin{tabular}{lcc}
\toprule
 & (1) & (2) \\
 & No controls & With controls \\
\midrule

\multicolumn{3}{l}{\textbf{Panel A: Log Weekly Working Hours}} \\
\midrule

Field premium $\times$ Covid & 0.010*** & 0.009*** \\
                             & (0.002) & (0.002) \\

\hline
Observations & 107,266 & 107,266 \\
R-squared & 0.108 & 0.147 \\

\midrule
\multicolumn{3}{l}{\textbf{Panel B: Log Monthly Earnings}} \\
\midrule

Field premium $\times$ Covid & 0.008 & 0.013** \\
                             & (0.007) & (0.006) \\

\hline
Observations & 99,137 & 99,137 \\
R-squared & 0.192 & 0.322 \\

\midrule
District FE & Yes & Yes \\
Year FE & Yes & Yes \\
Controls & No & Yes \\
\bottomrule
\end{tabular}

\begin{tablenotes}
\footnotesize
\item Estimates using eq.~(\ref{eq:main-eq}) where $Covid$ is time dummy for all COVID-period. All specifications include district and year fixed effects. Robust standard errors are reported in parentheses. *** $p<0.01$, ** $p<0.05$, * $p<0.1$.
\end{tablenotes}
\end{threeparttable}
\end{table}

\section{Concluding Remarks}\label{concl}
In this paper, I examine whether and to what extent field of study matters for resilient labor market outcomes during large negative shocks using the recent COVID-19 pandemic. I first construct reduced-form premium for major technical fields of study, with non-technical education as baseline category. Next, I use these field premia as continuous treatment in the difference-in-difference framework to study how these premia shaped labor market outcomes. The results show that higher premia is associated with both weekly working hours and monthly earnings during the pandemic period but such advantages did not evolve when the pandemic hit. Instead, workers in higher-premium fields experienced higher weekly working hours in the second and third pandemic phases, while earnings advantages became statistically significant only in the final phase.\\

These findings suggest that individuals in high-premium fields were better positioned to adapt during prolonged labor market disruptions. This indicates that lower-premium fields were more exposed to persistent shocks. This has important policy implications. Expanding access to high-return fields may improve long-run labor-market resilience, but it is unlikely to be sufficient during crises. Workers in lower-return fields may require targeted support, including income protection, retraining, and assistance with job transitions, to prevent temporary downturns from widening field-based inequalities. \\

A key drawback of this study is the lack of detailed information on fields of specialization in PLFS. This limits the construction of more granular major-specific premia. Nevertheless, these findings reflect not only the resilience of high-premium fields but also depict differential recovery trends across fields throughout the pandemic-induced downturns.

\section*{Data Statement}
The data has restricted access but free of cost. One needs to register with the \href{https://microdata.gov.in/NADA/index.php/home} {Microdata Archive} and comply with MoSPI data-use agreement. Once you sign up and download the data, you can replicate this analysis using the package: \url{https://doi.org/10.5281/zenodo.19546228}

\section*{Declaration of generative AI and AI-assisted technologies in the manuscript preparation process}
The author has used ChatGPT for text editing and language refinement. All content was reviewed, edited and verified by the author who takes full responsibility.

\section*{Declaration of competing interest}
The authors declare that they have no known competing financial interests or personal relationships that could have impact the work reported in this paper.

\section*{Funding Sources}
This research did not receive any specific grant from funding agencies in the public, commercial, or not-for-profit sectors.

\section*{Acknowledgments}
An earlier version of this paper was presented at the Graduate Student Seminar and Applied Microeconomics working group at the American University. I thank the participants for their helpful comments and suggestions. 

\newpage
\printbibliography

@techreport{NBERw29605,
 title = "College Majors and Skills: Evidence from the Universe of Online Job Ads",
 author = "Hemelt, Steven W and Hershbein, Brad and Martin, Shawn M and Stange, Kevin M",
 institution = "National Bureau of Economic Research",
 type = "Working Paper",
 series = "Working Paper Series",
 number = "29605",
 year = "2021",
 month = {12},
 doi = {10.3386/w29605},
 URL = "http://www.nber.org/papers/w29605",
}

@article{Weiss_why_2024,
 author = {Weiss, Deborah M. and Spitzer, Matthew L. and Cronin, Colton and Chin, Neil},
 issn = {1465-7287},
 journal = {Contemporary Economic Policy},
 Month = {1},
 number = {2},
 pages = {278--304},
 publisher = {Wiley},
 title = {{Why} college majors and selectivity matter: {Major} groupings, occupation specificity, and job skills},
 url = {http://dx.doi.org/10.1111/coep.12634},
 volume = {42},
 year = {2024}
}

@book{Callaway_difference-in-differences_2024,
 author = {Callaway, Brantly and Goodman-Bacon, Andrew and Sant’Anna, Pedro H.},
 institution = {National Bureau of Economic Research},
 month = {February},
 title = {{Difference-in-differences} with a {Continuous} {Treatment}},
 url = {http://dx.doi.org/10.3386/w32117},
 year = {2024}
}

@article{Andrews_the_2024,
 author = {Andrews, Rodney J. and Imberman, Scott A. and Lovenheim, Michael F. and Stange, Kevin},
 issn = {1530-9142},
 journal = {Review of Economics and Statistics},
 Month = {9},
 pages = {1--45},
 publisher = {MIT Press},
 title = {{The} {Returns} to {College} {Major} {Choice}: {Average} and {Distributional} {Effects}, {Career} {Trajectories}, and {Earnings} {Variability}},
 url = {http://dx.doi.org/10.1162/rest_a_01503},
 year = {2024}
}

@article{LeightonSpeer2020,
  author  = {Leighton, Mark and Speer, Jamin},
  title   = {The Role of Degree Specificity in Early Career Outcomes},
  journal = {Labour Economics},
  year    = {2020},
  volume  = {65},
  pages   = {101865},
  doi     = {10.1016/j.labeco.2020.101865}
}

@article{Davis_stem_2026,
 author = {Davis, James C. and Diethorn, Holden A. and Marschke, Gerald R. and Wang, Andrew J.},
 issn = {0048-7333},
 journal = {Research Policy},
 Month = {1},
 number = {1},
 pages = {105361},
 publisher = {Elsevier BV},
 title = {{STEM} employment resiliency during recessions: {Evidence} from the {COVID-19} {Pandemic}},
 url = {http://dx.doi.org/10.1016/j.respol.2025.105361},
 volume = {55},
 year = {2026}
}

@article{Adrjan_working_2025,
 author = {Adrjan, Pawel and Ciminelli, Gabriele and Judes, Alexandre and Koelle, Michael and Schwellnus, Cyrille and Sinclair, Tara M.},
 issn = {0927-5371},
 journal = {Labour Economics},
 month = {10},
 pages = {102751},
 publisher = {Elsevier BV},
 title = {{Working} from home after {COVID-19}: {Evidence} from job postings in 20 countries},
 url = {http://dx.doi.org/10.1016/j.labeco.2025.102751},
 volume = {96},
 year = {2025}
}

@article{Blanas_covid-induced_2023,
 author = {Blanas, Sotiris and Oikonomou, Rigas},
 issn = {0927-5371},
 journal = {Labour Economics},
 month = {4},
 pages = {102335},
 publisher = {Elsevier BV},
 title = {{COVID-induced} economic uncertainty, tasks and occupational demand.},
 url = {http://dx.doi.org/10.1016/j.labeco.2023.102335},
 volume = {81},
 year = {2023}
}

@article{Altonji_cashier_2016,
 author = {Altonji, Joseph G. and Kahn, Lisa B. and Speer, Jamin D.},
 issn = {1537-5307},
 journal = {Journal of Labor Economics},
 month = {1},
 number = {S1},
 pages = {S361--S401},
 publisher = {University of Chicago Press},
 title = {{Cashier} or {Consultant}? {Entry} {Labor} {Market} {Conditions}, {Field} of {Study}, and {Career} {Success}},
 url = {http://dx.doi.org/10.1086/682938},
 volume = {34},
 year = {2016}
}

@article{10.1257/aer.104.5.387,
Author = {Altonji, Joseph G. and Kahn, Lisa B. and Speer, Jamin D.},
Title = {Trends in Earnings Differentials across College Majors and the Changing Task Composition of Jobs},
Journal = {American Economic Review},
Volume = {104},
Number = {5},
Year = {2014},
Month = {5},
Pages = {387–93},
DOI = {10.1257/aer.104.5.387},
URL = {https://www.aeaweb.org/articles?id=10.1257/aer.104.5.387}}

@article{Walker_differences_2011,
 author = {Walker, Ian and Zhu, Yu},
 issn = {0272-7757},
 journal = {Economics of Education Review},
 Month = {12},
 number = {6},
 pages = {1177--1186},
 publisher = {Elsevier BV},
 title = {{Differences} by degree: {Evidence} of the net financial rates of return to undergraduate study for {England} and {Wales}},
 url = {http://dx.doi.org/10.1016/j.econedurev.2011.01.002},
 volume = {30},
 year = {2011}
}

@techreport{altonji2025returns,
  title = {Returns to Specific Graduate Degrees: Estimates Using Texas Administrative Records},
  author = {Altonji, Joseph and Zhu, Zhengren},
  institution = {National Bureau of Economic Research},
  type = {Working Paper},
  number = {33530},
  year = {2025},
  Month = {2},
  doi = {10.3386/w33530},
  url = {https://www.nber.org/papers/w33530}
}

@inbook{Altonji_the_2016,
 author = {Altonji, J.G. and Arcidiacono, P. and Maurel, A.},
 booktitle = {Handbook of the Economics of Education},
 isbn = {9780444634597},
 issn = {1574-0692},
 pages = {305--396},
 publisher = {Elsevier},
 title = {{The} {Analysis} of {Field} {Choice} in {College} and {Graduate} {School}},
 url = {http://dx.doi.org/10.1016/B978-0-444-63459-7.00007-5},
 year = {2016}
}

@article{Kahn_the_2010,
 author = {Kahn, Lisa B.},
 issn = {0927-5371},
 journal = {Labour Economics},
 month = {4},
 number = {2},
 pages = {303--316},
 publisher = {Elsevier BV},
 title = {{The} long-term labor market consequences of graduating from college in a bad economy},
 url = {http://dx.doi.org/10.1016/j.labeco.2009.09.002},
 volume = {17},
 year = {2010}
}

@article{Oreopoulos_the_2012,
 author = {Oreopoulos, Philip and von Wachter, Till and Heisz, Andrew},
 issn = {1945-7790},
 journal = {American Economic Journal: Applied Economics},
 Month = {1},
 number = {1},
 pages = {1--29},
 publisher = {American Economic Association},
 title = {{The} {Short-} and {Long-Term} {Career} {Effects} of {Graduating} in a {Recession}},
 url = {http://dx.doi.org/10.1257/app.4.1.1},
 volume = {4},
 year = {2012}
}

@article{Choi_the_2020,
 author = {Choi, Eleanor Jawon and Choi, Jaewoo and Son, Hyelim},
 issn = {0927-5371},
 journal = {Labour Economics},
 Month = {12},
 pages = {101926},
 publisher = {Elsevier BV},
 title = {{The} long-term effects of labor market entry in a recession: {Evidence} from the {Asian} financial crisis},
 url = {http://dx.doi.org/10.1016/j.labeco.2020.101926},
 volume = {67},
 year = {2020}
}

@article{Long_do_2015,
 author = {Long, Mark C. and Goldhaber, Dan and Huntington-Klein, Nick},
 issn = {0272-7757},
 journal = {Economics of Education Review},
 Month = {12},
 pages = {1--14},
 publisher = {Elsevier BV},
 title = {{Do} completed college majors respond to changes in wages?},
 url = {http://dx.doi.org/10.1016/j.econedurev.2015.07.007},
 volume = {49},
 year = {2015}
}

@article{yamada2018college,
  author  = {Yamada, Gustavo A. and Lavado, Pablo},
  title   = {Labor Market Consequences of the College Boom Around the World},
  journal = {IZA World of Labor},
  year    = {2018},
  number  = {165},
  doi     = {10.15185/izawol.165.v2},
  url     = {https://doi.org/10.15185/izawol.165.v2}
}

@book{StateOfWorkingIndia2026,
  title        = {Youth in the Labour Market: Pathways from Learning to Earning},
  author       = {{State of Working India}},
  year         = {2026},
  publisher    = {Centre for Sustainable Employment, Azim Premji University},
  url          = {https://publications.azimpremjiuniversity.edu.in/6848/1/SWI%202026%20-%20Web.pdf}
}

@article{Mehrotra2014,
  author  = {Mehrotra, Santosh and Parida, Jajati K. and Sinha, Suman and Gandhi, Ankita},
  title   = {Explaining Employment Trends in the Indian Economy: 1993--94 to 2011--12},
  journal = {Economic and Political Weekly},
  year    = {2014},
  volume  = {49},
  number  = {32},
  pages   = {49--57}
}

\newpage

\appendix
\renewcommand{\thesection}{A}
\renewcommand{\thesubsection}{A.\arabic{subsection}}
\renewcommand{\thefigure}{A.\arabic{figure}}
\setcounter{figure}{0}
\section* {Appendix}
\subsection{Descriptive Plots of Labor Outcomes}
\begin{figure}[h]
    \centering
    \captionsetup{width=0.95\linewidth}
    \includegraphics[width=\linewidth]{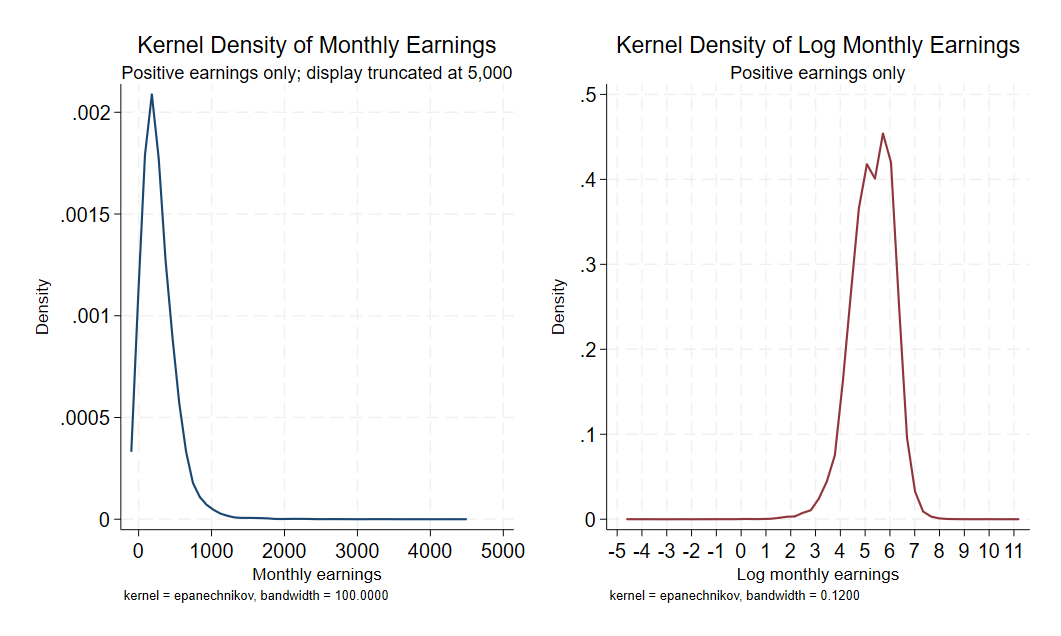}
    \caption{\textit{Kernel density of monthly earnings. The left panel shows the distribution in levels, which is strongly right-skewed. The right panel shows the distribution of log monthly earnings, which is more symmetric after the log transformation.}}
    \label{fig:fig_summ1}
\end{figure}

\begin{figure}[h]
    \centering
    \captionsetup{width=0.8\linewidth}
    \includegraphics[width=\linewidth]{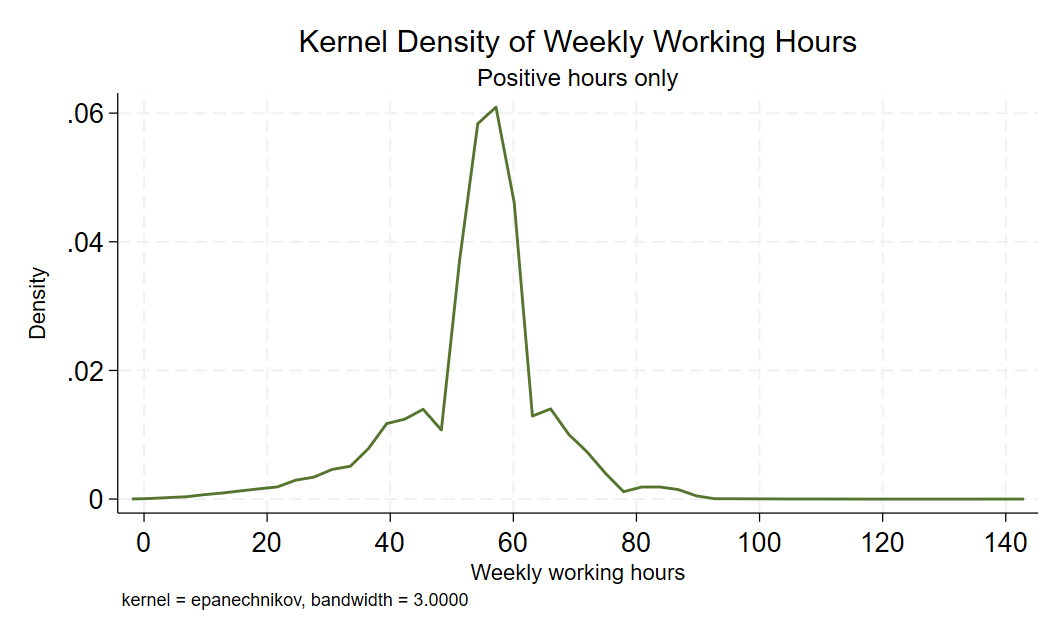}
    \caption{\textit{Kernel density of positive weekly working hours. The peak occurs near 50-60 hours per week while the tail becomes thinner at 100 hours per week and above.}}
    \label{fig:fig_summ2}
\end{figure}

\end{document}